\RequirePackage[hyphens]{url}

\documentclass[a4paper,fleqn]{cas-dc}
\usepackage{cas-dfrws} %overwrite some of the original (cas-dc.cls)
\renewcommand{\dfrwsconference}{Digital Forensics Research Conference Europe
(DFRWS EU) 2021,  March 29--April 1, 2024}

\usepackage[numbers,sort]{natbib}
\usepackage[english]{babel}
\usepackage[utf8x]{inputenc}
\usepackage{glossaries}
\setacronymstyle{long-short}

\newcommand{\cmd}[1]{\texttt{#1}} % For code or commands
\newacronym{emmc}{eMMC}{Embedded MultiMediaCard}
\newacronym{onfi}{ONFi}{Open NAND Flash Interface Specification}
\newacronym{jedec}{JEDEC}{Joint Electron Device Engineering Council}
\newacronym{mmca}{MMCA}{MultiMediaCard Association}

\newcommand{\FAU}{FAU}
\newcommand{\Leiden}{Leiden}
\newcommand{\AlbSig}{Albstadt}
\newcommand{\MSAB}{MSAB}

\begin{document}
\shorttitle{In Search of Lost Data}
\shortauthors{Schneider et~al.}

\title[mode = title]{In Search of Lost Data: A Study of Flash Sanitization Practices}

\author[1]{Janine Schneider}
\ead{janine.schneider@fau.de}
\credit{Conceptualization, Data curation, \\Formal Analysis, Investigation, Methodology, Software, Supervision, Validation, Visualization, Writing -- original draft, Writing -- review \& editing}
\author[1]{Immanuel Lautner}
\ead{immanuel.lautner@fau.de}
\credit{Data \\curation, Investigation, Validation, Writing -- original draft}
\author[1]{Denise Moussa}
\ead{denise.moussa@fau.de}
\credit{Data curation, Investigation, Validation, Writing -- original draft}
\author[1]{Julian Wolf}
\ead{julian.jw.wolf@fau.de}
\credit{Investigation, Writing -- original draft, Writing -- review \& editing}
\author[1]{Nicole Scheler}
\ead{nicole.scheler@fau.de}
\credit{Writing -- original draft, Writing -- review \& editing}
\author[1]{Felix Freiling}
\cormark[1]
\ead{felix.freiling@cs.fau.de}
\credit{Conceptualization, Methodology, Supervision, Writing -- original draft, Writing -- review \& editing}
\author[2]{Jaap Haasnoot}
\ead{haasnoot.j@hsleiden.nl}
\credit{Data curation, Investigation}
\author[2]{Hans Henseler}
\ead{henseler.h@hsleiden.nl}
\credit{Conceptualization}
\author[3]{Simon Malik}
\ead{maliks@hs-albsig.de}
\credit{Data curation, Investigation}
\author[3]{Holger Morgenstern}
\ead{morgenstern@hs-albsig.de}
\credit{Conceptualization}
\author[4]{Martin Westman}
\ead{Martin.Westman@msab.com}
\credit{Conceptualization, Resources}
\address[1]{Department of Computer Science,
Friedrich-Alexander-Universit\"at Erlangen-N\"urnberg (FAU),
Erlangen, Germany}
\address[2]{Leiden University of Applied Sciences, Leiden, The Netherlands}
\address[3]{Albstadt-Sigmaringen University, Albstadt, Germany}
\address[4]{Micro Systemation (MSAB), Stockholm, Sweden}
\cortext[cor1]{Corresponding author}
	
\begin{abstract}
	To avoid the disclosure of personal or corporate data, sanitization
	of storage devices is an important issue when such devices are to be
	reused. While poor sanitization practices have been reported for
	\emph{second-hand} hard disk drives, it has been reported that data
	has been found on \emph{original} storage devices based on flash
	technology.  Based on insights into the second-hand chip market in
	China, we report on the results of the first large-scale study on
	the effects of chip reuse for USB flash drives. We provide clear
	evidence of poor sanitization practices in a non-negligible fraction
	of USB flash drives from the low-cost Chinese market that were sold
	as original.  More specifically, we forensically analyzed 614 USB
	flash drives and were able to recover non-trivial user data on a
	total of 75 devices (more than 12 \%).  This non-negligible
	probability that any data (including incriminating files) already
	existed on the drive when it was bought has critical implications to
	forensic investigations. The absence of external factors which
	correlate with finding data on new USB flash drives complicates the
	matter further.
\end{abstract}


\maketitle

\nocite{Proust:2001:ISL}% hint to the title

% Note: Garfinkel's article starts its title with "Remembrance of data
% passed" which appears to be an allusion to "Remembrance of things
% passed" which in turn is the (former) title of the English
% translation of Marcel Proust's monumental novel "a la recherche du
% temps perdu", now translated as "In Search of Lost Time". Likewise
% "Remembrance of things passed" is an allusion to Shakespeare's
% Sonnet 30
%
% https://en.wikipedia.org/wiki/Sonnet_30
% https://en.wikipedia.org/wiki/In_Search_of_Lost_Time

\section{Introduction}
\label{sec:intro}

The increase of storage capacity of hard disk drives and the
corresponding decrease in costs over the last 20 years has paved the
way for a relevant second-hand market for storage technology. As
Garfinkel and Shelat
\cite{DBLP:journals/ieeesp/GarfinkelS03,DBLP:journals/dsonline/GarfinkelS03}
showed many years ago, such devices often ``contain information that
is both confidential and recoverable'': Of the 83 second-hand hard
disk drives they acquired in 2003, a total of 49 contained recoverable
data including credit-card information, corporate memoranda and
personal medical records. This was corroborated later by Freiling et
al.~\cite{DBLP:conf/imf/FreilingHM08} who reported on similar and even
more privacy-invading results.
% , e.g., the reconstruction of the
% complete family tree of the former owner of a hard disk in a forensic
% exercise with University students.

Today, it is well-known that data can be recovered from disk drives
unless effort is spent on its deletion
\cite{DBLP:conf/iciss/WrightKS08}, and poor sanitization practices
have even reached popular culture like art expositions
\cite{Heise:2019}. Given the rather aggressive way in which solid
state drives reclaim deleted data \cite{Nisbet:2013:AFA}, sanitization
appears to have become easier with storage based on NAND-flash
technology. 
This hypothesis is further supported by the fact that even forensic scientists encounter problems when examining NAND-based storage.
The forensic analysis of NAND-based storage has already been discussed extensively \cite{10.1007/978-3-030-05487-8_8, FUKAMI2017S1, 10.1007/978-3-319-46279-0_18, Reddy2019}.
For USB flash drives and memory cards sold on the
second-hand market poor sanitization practices have been confirmed by
Robins et al.~\cite{doi:10.1080/1206212X.2017.1289689} and numerous
other popular studies. So overall it is clear that nobody today can
safely assume that a second-hand storage device does \emph{not}
contain any data from previous use.

% today: An art project with recovered images from used disks: note on Heise on 9.8.2019 \cite{Heise:2019} with reference to the art gallery and exhibition in Stuttgart \url{https://www.galerie-kernweine.com/#programm}

From the viewpoint of a forensic investigator, poor sanitization
practices are relevant if incriminating data is found on a storage
device claimed to have been obtained second hand. In contrast, buying a
new USB flash thumb drive was usually considered safe because no previous
use implies no previous data. It was therefore rather surprising when
Westman \cite{Westman:2018:WDT,FAZ:2017} reported that he had found
non-trivial data on \emph{new} USB drives.

It has been speculated that Westman's findings were due to the reuse
of memory chips in USB devices. Naturally, reused memory chips are
much cheaper than new ones and can be bought on specialized markets.
These circumstances and the possibility of finding old data on new
drives have caused great concern in the digital forensic community
since attribution of data found on new devices becomes as difficult as
for second-hand ones. It is therefore important to assess and somehow
quantify the risks of acquiring evidence of former usage on such new
USB drives.

% \begin{work}
% %  Maybe mention related work on chip recycling?
% %   % 
% % 	\begin{itemize}
% % 		\item Bachelor project: Chip Recycling: Recycling of Chips from BZZ
% % 		Conditioning Processes
% % 		(\url{https://web.wpi.edu/Pubs/E-project/.../E.../YR_SGMQP_05.pdf},
% % 		in papers subdirectory)
% % % https://web.wpi.edu/Pubs/E-project/Available/E-project-042607-151242/unrestricted/YR_SGMQP_05.pdf
                
% % 		\item Contacts: Martin Westmann, MSAB
% % 		\item Project ``faire computer'' by FIfF, see
% % 		\url{https://www.fiff.de/themen/fair_it}, contact Sebastian
% % 		Jekutsch
% % 	\end{itemize}
% \end{work}

% This document lays out a plan for an experimental study to investigate
% this issue.

% We first lay out the research question and then give some background
% on existing knowledge about the production of USB drives. Next, we
% describe the study design and working instructions for those
% researchers participating in the study.

\subsection{Research Questions}
\label{sec:research:questions}

Based on reports on chip recycling in the
literature~\cite{wang2008economic,chi2011informal}
we investigate the following research question:
\begin{quote}
  What is the risk of acquiring evidence on new USB drives that are
  due to former usage (of components) of the drive, i.e., usage before
  it was bought?
\end{quote}
Surely, the percentage of USB flash drives that contain a recycled
memory chip would be an upper bound on the probability to find such
data, but we are not aware of any data that quantifies this
percentage. As we describe later in this paper, manufacturers are also
naturally reluctant to disclose the fact that second-hand components
are built-in to new devices. Therefore, we have to examine a
sufficiently large sample of ``new'' flash drives to approximate the
risk of finding data.  On such drives, we take the existence of
non-trivial (user) data on a new USB drive as the only measurable and
clear indicator of memory chip recycling. 
%    
% In contrast, the \emph{non-existence} of such data does not
% necessarily mean that the chip is \emph{not} recycled.
%

Obviously, the existence of such non-trivial data is a critical issue
of forensic concern, since a suspect may now claim in many
circumstances that incriminating material had been on the drive when
he bought the device. To distinguish this from the case that the
suspect himself had planted the data on the disk, ideally, a new drive
would have a label ``contains traces of former usage'' when it was
bought. In case of a dispute, the label would allow an investigator to
estimate the probability that the suspect himself had stored the file
on the drive. But since such a label unfortunately does not exist, we
ask ourselves whether there are any indicators that can stand-in for
this information, e.g., visual appearance, capacity,
built-in-technology, manufacturer, or manufacturing date. So, our
second research question is:
\begin{quote}
  What factors influence the probability for the existence of
  non-trivial user data on new USB drives?
\end{quote}
As we assume that the probability of getting a recycled memory chip
within a USB drive is inversely proportional to the price, we try to
maximize this probability by focusing on the low-cost market of
promotional USB drive products.  For such cheap USB drives, our goal
is therefore to measure (1) the probability of occurrence of old data
and (2) the quantitative correlation between certain external factors
of the USB drive and the existence of non-trivial user data on USB
drives.

\subsection{Contributions}
\label{sec:contributions}

To summarize, the contributions of this paper are as follows:
\begin{itemize}
\item We report on the results of the first large-scale study on the
  effects of chip reuse for cheap/promotional USB flash drives.
\item We provide clear evidence of poor sanitization practices in a
  non-negligible fraction of USB flash drives from the low-cost
  Chinese market that were sold as original.
\item More specifically, we forensically analyzed 614 USB flash drives
  and were able to recover non-trivial user data on a total of 75
  devices (more than 12 \%).
\item Apart from finding data on the device, we found no other clear
  predicting indicator of chip reuse neither through external factors nor
  through internal ones.
\item We discuss the methodological and legal consequences of these
  findings to forensic investigations.
\end{itemize}

\subsection{Paper Outline}

We provide some background on NAND flash and relevant technologies in
Section~\ref{sec:background} and on the flash drive market in
Section~\ref{sec:flash:market}. We then present the design of our
study and the results in Sections~\ref{sec:study:design} and
\ref{sec:results}. Legal implications are discussed in
Section~\ref{sec:legal} before we conclude in
Section~\ref{sec:summary}.

\section{Background}
\label{sec:background}

Today USB thumb drives are mainly built from memory chips that have a
built-in controller circuit. Most of these chips follow the \gls{emmc}
standard while some also adhere to the \gls{onfi}. As background
information, we therefore briefly give an overview over the involved
technologies with focus on their possibilities for disk sanitization.

\subsection{Flash Storage}
NAND flash storage is a non-volatile storage consisting of transistors using the floating gate technology \cite{SSD}.
These transistors are able to push electrons onto an electrically insulated gate, which are then trapped and remain there even if all applied voltages are removed.
The charge of the gate causes an increase of the threshold voltage, at which the transistor becomes conductive on the source-drain path.
These processes can be digitally evaluated, where introducing electrons on the floating gate causes a logical 0 and removing electrons from the floating gate causes a logical 1.
Furthermore, by a controlled intrusion of charges into the isolated gate, several changing states of the threshold voltage can be generated and read out again, with the effect that several bits can be stored in these cells.

How many bits can be stored per cell depends on the NAND technology
and can range from Single-level cells \\(SLC, 1 bit per cell) over
Multi-level cells (MLC, 2 bits) up to Triple-level cells (TLC, 3
bits).

\subsection{eMMC Technology}

The \gls{emmc} standard was introduced in November 2007 by the
\gls{jedec} and the \gls{mmca} in order to provide a data storage and
communication media for a great number of mobile devices. The
technology aims at meeting the performance requirements of such
devices while keeping power consumption low~\cite{JESD84-A43:2007}.
An \gls{emmc} media is closely related to a Multi Media Card (MMC),
since it is a managed NAND flash package, where the MMC components,
flash media and device controller are in one unit. Data transfers
happen via a configurable number of data bus signals. For issuing
commands, a bidirectional command channel signal is defined by the
standard.

There are 64 different commands (CMD0--CMD63) with fixed length of 48
bits, where some can take a 32 bit argument, often a memory
address. Commands are sent from a host controller to the \gls{emmc}
device, whereas responses are sent back from device to host
controller~\cite{JESD84-B51:2015}. In the \gls{emmc} standard, host
addresses can either be \emph{mapped} or \emph{unmapped}. The mapped
host address range defines the addresses of the \gls{emmc} device that
can be accessed by a read command from the host software.

% \paragraph{Data Removal Operations.}

An erasable unit of a \gls{emmc} is called `Erase Group' and consists
of a device specific number of write blocks which are the basic
writeable units.  An erase process is a three step sequence,
consisting of the following commands:
\begin{enumerate}
\item ERASE\_GROUP\_START (CMD35) defines the start address of the
  range to be erased,
\item ERASE\_GROUP\_END (CMD36) defines the end address of the range, and
\item ERASE (CMD38) starts the actual removal process.
\end{enumerate}
Depending on the argument given to CMD38, six different erasure
behaviors were defined by the standard until today:
\begin{itemize}
\item The most basic form of \emph{Erase} was already available since version 4.3 in 2007~\cite{JESD84-A43:2007}.
\item In 2009 (version 4.4) \emph{TRIM}, \emph{Secure TRIM Step 1},
  \emph{Secure TRIM Step 2} and \emph{Secure Erase} were added.
\item Since 2011 (version 4.5) the erasure behavior \emph{Discard} is
  possible \cite{JESD84-B51:2015}.
\end{itemize}

\subsubsection{Simple Data Removal Operations}
\label{sec:simpledr}

The \emph{Erase} behavior will eventually erase the specified groups,
but the controller is not forced to perform physical erasure at this
point but can schedule it to a convenient time. Similarly, \emph{TRIM}
results in an application of the erase operation but not on erase
groups but write blocks. The host can flag no longer required blocks
for erasure so that the device can erase them during background erase
events. Partial or full actual erasure of the flagged blocks is again
performed at a convenient time by the controller.

When an \emph{Erase} or \emph{TRIM} command finishes with success, the
targeted device address range behaves as if it was overwritten
completely with zeros or ones (depending on the technology) and the
specified address range moves to the unmapped host address range.

% TODO: Absatz überflüssig?!
%	For erasing an individual write block the TRIM command is mostly the better choice, since the Erase command eventually erases all blocks within an erase group and still needed data may get lost. 
	
\emph{Discard} is an operation similar to \emph{TRIM}, the difference
being that a discarded region may return parts or all of the original
data. Parts that do not return data anymore behave like trimmed or
erased parts, while data areas eventually should be moved to the
unmapped address region. The controller can but does not have to
perform full or partial erasure~\cite{JESD84-B51:2015}.

\subsubsection{Secure Data Removal}
\label{sec:securedr}

Since the above erasure behaviors leave a lot of choices for the
device whether and when to delete data, version 4.4 of the \gls{emmc}
standard introduced \emph{Secure Erase} and \emph{Secure TRIM}.  The
difference between \emph{Erase} and \emph{Secure Erase} is that the
latter blocks any other command to be processed by the device until
the actual erasure is completed. Furthermore, a \emph{secure purge}
operation is performed on the erase groups and on any copies of items
in those erase groups. On success, the operation results in removing
all data from the unmapped host address space.
%
% What the secure purge operation does in detail is device specific as
% it can be configured by setting the SECURE\_REMOVAL\_TYPE field in the
% Extended Device Specific Data (EXT\_CSD) register of the \gls{emmc}
% unit. The register (introduced in version 4.0) stores information
% about device capabilities and selected modes. The standard specifies
% four different possible options for information removal:
% %
% \begin{enumerate}
% \item erase of physical memory (actual NAND erase)
% \item overwriting the addressed locations with a character followed by an erase
% \item overwriting the addressed locations with a character, its complement, then a random character
% \item using a vendor defined procedure
% \end{enumerate}
%
Similarly, \emph{Secure TRIM} differs from \emph{TRIM} by performing a
secure purge operation on write blocks.

To minimize impacts on performance, \emph{Secure TRIM} is divided in
two steps: \emph{Secure TRIM Step 1} and \emph{Secure TRIM Step 2}.
With the first step, a range of write blocks can be marked for the
secure purge operation using CMD35 and CMD36. The second step then
performs the actual secure removal the same way \emph{Secure Erase}
does. The two commands themselves cannot be interrupted but it is
possible to issue commands between them. Note that if a block marked
for erasure is written before \emph{Secure TRIM Step 2} is issued,
then this last copy will not be marked and therefore will remain
untouched by the erase operation.

The above set of data removal operations were deprecated in versions
later than 4.51 in favor of using an \emph{Erase}/\\\emph{TRIM} followed
by a \emph{Sanitize} operation to achieve the same result.
\emph{Sanitize} (which was actually added in version 4.5) forces the
device to remove all data from the unmapped user address
space~\cite{JESD84-B51:2015}. 

% It is initiated by writing a value to the Device Specific Data
% register (CSD) of the \gls{emmc}, which stores information about the
% device operation conditions~\cite{JESD84-B51:2015}.

% \paragraph{Discussion.}

% Simple Data Removal \ref{sec:simpledr} does not necessarily delete the
% targeted data physically.  The \emph{Discard} operation does not
% guarantee that the targeted data is actually erased from memory while
% the \emph{Erase} and \emph{TRIM} operation allow the controller to
% schedule the actual erasure for later but the \gls{emmc} standard does
% not specify that point in time any further. To account for the need of
% deleting data directly after the command is issued, \emph{secure}
% erasure processes \ref{sec:securedr} are provided by the \gls{emmc}
% standard. All processes are terminated by a hardware reset or a power
% failure leaving the data in an unknown state~\cite{JESD84-B51:2015}.

\subsection{ONFi Technology}

ONFi is an industry workgroup made up of more than 100 companies
defining standardized component-level interface specifications as well
as connector and module form factor specifications for NAND
Flash. Their aim is to increase compatibility and interoperability of
NAND devices from different vendors. Contrary to \gls{emmc}, ONFi is
no card standard because it only defines the interface to the NAND
flash component itself but excludes the specification of a device
controller. The first ONFi specification (ONFi 1.0) was released in
December 2006. The latest version 4.2 is from February 2020 (see
\url{www.onfi.org}).
	 
ONFi defines the device as a packaged NAND unit. It consists of one
ore more NAND targets which are made of an arbitrary number of logical
unit numbers (LUNs). Each LUN can execute commands and report status
independently. A LUN consists of an arbitrary number of blocks. A
block contains a number of pages and is the smallest erasable
unit. Each page optionally consists of partial pages which are the
smallest unit to program or read~\cite{ONFI4.2:2020}.
	 
A LUN generally can perform physical erase operations on
blocks. Depending on the device controller, which is not part of the
ONFi specification, different erase routines may be defined by
manufacturers. In contrast to the \gls{emmc} standard, it is not
possible to generally describe what data removal processes can be
expected for ONFi devices.

\section{The USB Flash Drive Market}
\label{sec:flash:market}

USB flash drives are an increasingly important market sector for cheap
end user storage devices. We now take a look at the players in the
market of promotional products and the economic incentives that
motivate chip recycling in this industry.

\subsection{Market Players}
\label{sec:market_players}

According to the International Network of the Promotional Product
Industry \cite{PSI}, there are five basic roles in the industry of
promotional products, including give-away USB drives:
\begin{enumerate}
\item \emph{Manufacturers} % (Hersteller)
  are companies that assemble electronics and the packaging of a USB drive.
\item \emph{Promotional product suppliers} % (Lieferant)
  can be a manufacturer or merely import products from another manufacturer.
\item \emph{Distributors} % (H\"andler)
  are companies that source promotional products.
\item A \emph{finisher} % (Veredler):
  takes the basic product from a distributor or supplier and prints,
  lasers, engraves, paints, stamps, etc.~the product to bring it into
  shape.
\item Finally, an \emph{advertising} or \emph{media
    agency} % (Werbe-Agentur):
  gives full service to customers as part of public relations
  campaigns. This includes the acquisition of promotional products
  from finishers.
\end{enumerate}
In order to receive promotional USB drives, customers from Europe
usually contact an agency, a distributor or a supplier.
%
% For example, \FAU{} gets their promotional USB drives from a
% distributor in Germany that serves many universities. That company has
% contacts to several suppliers who collaborate with manufacturers in
% China that source their eMMC chips usually from Taiwan or South Korea.
%
A major fraction of manufacturers for USB thumb drives come from China
with the Chinese manufacturing industry having its center in Shenzhen,
Huangdong region, in southern China. These manufacturers usually
source their eMMC chips from Taiwan or South Korea.

% \item Research on the industrial ecosystem in southern China can be
%   supported by Auslandshandelskammer S\"udchine (\url{china.ahk.de}).

% Regarding USB/powerband products, there are about 90 suppliers in
% Germany and maybe 12 specialized in USB drives.
%
% German customers usually want good quality, so it is hard to get really
% cheap drives through the common agencies or distributors in Germany.
%

Distributors usually differentiate between class A and class B
drives. % (``A-Ware'', ``B-Ware'').
Class A drives contain new and fully functional memory chips. Class B
drives can contain memory chips that might not have the full capacity
because of malfunctions on the chip or production problems.  While it
is rather unusual to be able to explicitly order class B drives, it is
highly probably to get such drives when selecting by price and buy the
``cheapest of the cheapest'' ones. It is possible to also order USB
drives from manufacturers directly via portals like
Alibaba (\url{www.alibaba.com}).

% \item website of Hongkong Trade \& Development Council (because
%   many distributors and based in Hongkong): They have a platform
%   to research manufacturers of USB drives
% \end{enumerate}

% ANON: add footnote in final paper with this trivia
% \item Trivia: In Germany, you have to pay a form of import tax
%   (GEMA-Geb"uhr) on USB drives (0,91 EUR for up to 4 GB, and 1,56
%   EUR for larger drives).

\subsection{The Chinese Chip Recycling Market}
\label{sec:recycling_market}

%  Main contributor: Immanuel Lautner

For years China was the most important importer of e-waste. This
changed only because of the import ban on several kinds of waste in
2018 and its extension in 2019. Additionally, in the last years the
amount of e-waste that is produced inside China itself has grown every
year. This results in the need for recycling and remanufacturing of
e-waste, a need which has been picked up by the informal sector of
e-waste recycling, a sector that bears a relevant socio-economic role
in some cities in China \cite{wang2008economic,chi2011informal}.

The Chinese government has tried to regulate the informal sector by
facilitating a formal sector. However, this could not eliminate the
informal sector successfully, in particular because the informal
sector has a functioning network of individual collectors which are
the preferred recycling option for the Chinese households. After the
collectors gather the e-waste, it gets distributed by e-waste traders
to local informal recyclers. As the economic disparity is high, the
workers live from a low income which allows the process to be
cost-effective. The recyclers use manual and low-tech techniques to
dismantle the waste and extract valuable components, such as using
tools like hammers or heating the waste with coal-fired ovens.  As the
revenue is higher, the reuse of \emph{components} from e-waste (like
memory chips) is preferred in comparison to simply selling the raw
material like copper. Additionally, if there is the possibility to
reuse and resell an appliance directly this is preferred
\cite{chi2011informal}.

After the material is recycled, it gets resold to the respective
market. Given the informal nature of the recycling process, it is
difficult to determine whether a chip is new or definitely
remanufactured. But with the low-costs and the barely existing
regulations, there are strong incentives to declare old as new. At
eMMC spot markets such as \url{DRAMexhange.com} or
\url{en.chinaflashmarket.com} a 64 GB eMMC chip cost around 7 USD.

% some prices as of March 2018, all for eMMC:
% 	16 GB MLC: 6,44 USD
% 	8 GB MLC: 3,85 USD
% 	4 GB MLC: 3,02 USD
% 	32 GB MLC v.4.5/v.5.0: 11,20 USD
% 	8 GB TLC v.4.5/v.5.0: 3,75 USD
% 	16 GB TLC v.4.5/v.5.0: 5,40 USD

In the current state of the Chinese market, there are two
possibilities how a reused chip can find its way into a new USB
drive. Firstly, the USB drive could be reused as a whole. Secondly, a
different appliances (e.g., a mobile-phone or a smart TV) gets
recycled and the memory chip is removed and resold to a manufacturer.
Here it could be profitable to mix old and new chips to further
conceal the provenance. The manufacturer, knowingly or unknowingly,
uses this chip to manufacture a ``new'' USB drive. Given that a new
product first and foremost has to appear new, the second possibility
seems to be the more favorable one.

% \begin{work}
%   Discussion about the market of reused memory chips, e.g. from
%   Martin's introduction to the workshop in Florence
%   \cite{Westman:2018:WDT}:
%   % 
%   \item eMMC chips have ``stamps'' on them from quality control
%   \end{itemize}
% \end{work}

\section{Study Design}
\label{sec:study:design}

% \begin{work}
%   Main contributor: Julian Wolf
% \end{work}

We now describe the design of our study and the measurements we made.

\subsection{Possible Influencing Factors}
\label{sec:factors}

According to the first research question, the dependent variable we
wanted to predict was the probability of finding non-trivial data on a
new USB drive. After some initial research and before starting our
experiment, we collected the following set of independent variables
which we could measure and which potentially could have an effect on
the dependent variable:
\begin{itemize}
\item Manufacturer, location, company size, company name,
\item used NAND technology (SLC, MLC, TLC),
\item physical appearance (especially after opening the casing), since Westman had reported on potential signs of re-use like ``stamps'' on the eMMC chip \cite{Westman:2018:WDT},
\item size/capacity of eMMC chip in GB,
\item technology standard, for example the JEDEC standard for eMMC chips,
\item eMMC chip manufacturer (Samsung, Toshiba, etc.), and
\item cost.
\end{itemize}

\subsection{USB drive acquisition}

Due to the relatively large number of factors that could
potentially have an influence on the results, while at the same time
having little to no control over those factors when ordering those in
small batches from suppliers, the decision was made to consider two
primary factors when ordering the USB drives: cost and capacity. The
hypothesis for this experiment was that cheap and low quality drives
would have a higher chance of including recycled chips. Therefore it
was attempted to get very cheap drives directly from vendors selling
those via Alibaba.

Due to several research groups being involved in this project, the
acquisition was also done in a decentralized way, meaning each group
acquired their own drives for analysis. 
For the sake of simplicity, the participating research groups from 
Friedrich-Alexander-Universit\"at \\Erlangen-N\"urnberg, 
Leiden University of Applied Sciences, 
Albstadt-Sigmaringen University and
Micro Systemation
are referred to \FAU{}, \Leiden{}, \AlbSig{}, \MSAB{} in the following.
Group \FAU{} ordered a total
of 500 drives from 10 different suppliers, 50 drives per supplier,
which was usually considered the minimum order amount by vendors, and
also allows to analyze both variety across vendors as well as within a
vendor's batch. Five batches were ordered with 4 GB per drive, the
other five batches were ordered with 2 GB per drive to also have variety
in regards to capacity. The drives ordered by Group \FAU{} were ordered in
August and September 2018 and arrived between September and November
2018. The prices for 2 GB drives were between 2.08 USD and 3.50 USD
per drive, the prices for 4 GB drives ranged from 2.00 to 4.00 USD per
drive. The vendors were selected randomly from those selling USB flash
drives on Alibaba with the desired product portfolio (in regards to
price and capacity). Vendors were also not informed about the intended
analysis and usage for these drives. Group \FAU{} generously provided a
budget to buy these drives from funds to support teaching.

Group \Leiden{} acquired three batches of 4 GB drives \\through Alibaba
as well, following the same criteria as Group \FAU{}.  One smaller batch of
16 GB drives was provided by \MSAB{} for analysis and procured through
their own channels. Of those drives, 14 were analyzed at Group \Leiden{} and
16 at Group \FAU{}.

Group \AlbSig{} also acquired three batches of 2 GB and one batch of 16 GB drives (a total of 600 drives) through Alibaba.
The drives ordered by \AlbSig{} were ordered in November and December 2018 and arrived between December 2018 and January 2019.
Unfortunately, due to organizational complications and work overload the analysis of the USB drives was only partially completed by December 2020. To prevent the introduction of any bias caused by the selection of analyzed drives, we decided to not include the data from \AlbSig{} in our study.

% \begin{work}
%Add acquisition from ALBSIG, OSLO, LEIDEN. It appears:
%\\ALBSIG:
%\begin{itemize}
%\item 200x 2 GB
%\item 200x 2 GB
%\item 100x 16 GB
%\item One batch w/o info?
%\end{itemize}
%Oslo
%\begin{itemize}
%\item 500x 16 GB
%\end{itemize}
%Leiden
%\begin{itemize}
%\item 7 bayches, 3x 50x 2 GB, 4x unknown?
%\end{itemize}
%
%Need confirmation/verification
%\end{work}
	
\subsection{Measurements}
\label{sec:measurements}
	
Following the second research question, each influencing factor is an
independent variable while the dependent variable is the fact whether
we can find non-trivial user data on the USB drive. Data is
non-trivial user data if it is clearly distinguishable from random
data, e.g., viewable as an identifiable photographic image. For
simplicity, we refer to any non-trivial data as \emph{user data} if it
is found by a common file carver (like foremost \cite{foremost}) and
(after visual inspection) is not a false positive.

In preparation of the USB drive analysis, every supplier received an
ID, then every drive from that supplier's batch was labeled with the
supplier ID as well as a unique ID within that batch. This way, every
drive was uniquely labeled and could be traced back to the
supplier. The drives then were handed out to analysts with the task to
create a forensic 1:1 image, collect data, and analyze the data on
that image. The handling and analysis procedures were distributed in
writing and contained a list of well-defined and repeatable
acquisition and analysis steps.

The task required every analyst to document the following points per drive:
\begin{itemize}
\item Drive ID, supplier ID, analyst pseudonym, date imaged,
\item drive size in GB,
\item image SHA256 hash,
\item image MD5 hash (only for crossover validation),
\item NAND technology, eMMC standard, eMMC chip manufacturer,
\item distinctive visual features of the chip (after removing the packaging),
\item data found (yes/no), and 
\item if data was found: category of data found.
\end{itemize}
Not all data could be documented in all cases. For example, for many
drives the NAND technology, eMMC standard and eMMC chip manufacturer
could not be identified due to insufficient cues on the chip.

Most of the analysts were students, whereby each student received several drives to perform the
analysis steps.  Part of the acquisition and analysis procedures was
to protect any personal data found by encrypting it with a public key
to which only the instructors had access. Students had to sign a
declaration that they would keep any user data they found on the
drives confidential.

All results were double-checked by senior researchers before they
entered the dataset.  All USB drive images were stored on
access-restricted file storage and the dataset was stored in a central
database from which the results were computed.

% Additionally, 16 drives acquired by MSAB were analyzed as part of this project.
% The team at HS Leiden analyzed a total of 134 drives. 14 of those were provided by MSAB.

%\section{Working Instructions}
%\label{sec:instructions}
%	
%\begin{work}
%	We should take a random selection of manufacturers and suppliers
%	from the sources above.
%	
%	From each we should order a small set of USB drives. Go for the
%	cheapest possible. Usu
%	
%	Martin says: ``Say you want to buy usb sticks for commercial pr
%	material.  Buy a small batch off different sizes, just specify you
%	want emmc to be sure to get the one we get.  Or also include ``any''
%	to see how much you can interpret from the TSOPs as well.''
%	
%	What is a ``small batch''?
%\end{work}
%	
%\begin{work}
%	After receiving the drives:
%	
%	\begin{enumerate}
%		\item assign a unique ID to each one and label them
%		\item enter the factors together with the ID in the database
%		\item use dd or any other imaging tool to produce a 1:1 copy of the
%		drive to a file.
%		\item name the file \texttt{usb-<uniqueID>.dd}
%		\item Hash the file with SHA-256 and put the hash into the database.
%		\item Run a file carver (which one) on the file and document the
%		results in the database (see above).
%		\item Compress the file using gzip, ZIP or WinZip.
%		\item Send the zipped file to Felix.
%	\end{enumerate}
%\end{work}

\section{Results}
\label{sec:results}

We now report on the results of our study.

\subsection{General Information}

For the analysis of the experimental results, the datasets of the two
research groups were joined together.  As can be seen in
Fig.~\ref{tab:usbsticks} the joined dataset contained 650 USB drives
of which 614 finally evaluated.  36 USB drives were
sorted out because either the drive or the image was missing, the hash
sum was incorrect, or the drive did not work
anymore.
% , the drive could not be identified or other uncertainties have occurred.

Overall, 75 USB drives contained non-trivial user data to varying
degrees. However, any partially or fully reconstructable and readable
user data was considered. Random data, false positives from file
carving and corrupted data was sorted out. Results were generated using
carving tools like scalpel \cite{DBLP:conf/dfrws/RichardR05},
foremost \cite{foremost} and PhotoRec \cite{photorec}, but also commercial tools were used. 
The carvers were configured to maximize the amount of results, meaning that all search patterns were activated.
Furthermore, the analyst mostly used more than one carving tool.
In some cases, additional examinations like file system analysis was
performed.

Overall, we assume that the existence of non-trivial user data is an indication of chip recycling.
However, two USB drives (ID S5-41, S5-13) contained an active FAT32 file system including deleted FAT entries
and deleted private pictures. Interestingly, both USB drives contained
the very same set of pictures which were created and immediately
deleted shortly before the USB drives were shipped.  We therefore
assume that the pictures were used to test the functionality of the
USB drive and that the pictures are not artifacts of chip recycling.
Due to that, the data on these two USB drives was not considered to result from the use of recycled chips.
In total we therefore assume that recycled chips were used in at least 73 out of the 650 USB drives.

\begin{table}
  \begin{tabular}{|c|c|c|c|}
    \hline
    & \textbf{Group \FAU{}} & \textbf{Group \Leiden{}} & \textbf{Total} \\
    \hline
    \textbf{Total} & 516 & 134 & 650 \\
    \hline
    \textbf{Analyzed} & 484 & 130 & 614 \\
    \hline
    \textbf{Data found} & 61 & 14 & 75 \\
    \hline
  \end{tabular}
  \caption{Number of USB drives per research group. In total 614 out of 650 USB drives were analyzed, whereby 75 contained data.}
  \label{tab:usbsticks}
\end{table}

\subsection{Types of Data Found}

Carving the USB drives resulted in a huge amount of data.  Besides
many false positives, numerous different relevant data items could be
reconstructed.  The type of user data hereby varies considerably
between the different USB drives. Overall, the USB drives mainly
contained the following types of data:
\begin{itemize}
\item Gifs, icons, emojis and logos
\item Photos, pictures, wallpapers, maps
\item Music, film and series covers and posters
\item Ringtones
\item RPM, TAR and ZIP archives
\item Music, Videos and Movies
\item Speech recordings
\item Documents
\item Source code
\end{itemize}

\begin{figure}[h!]
  \includegraphics[width=\linewidth]{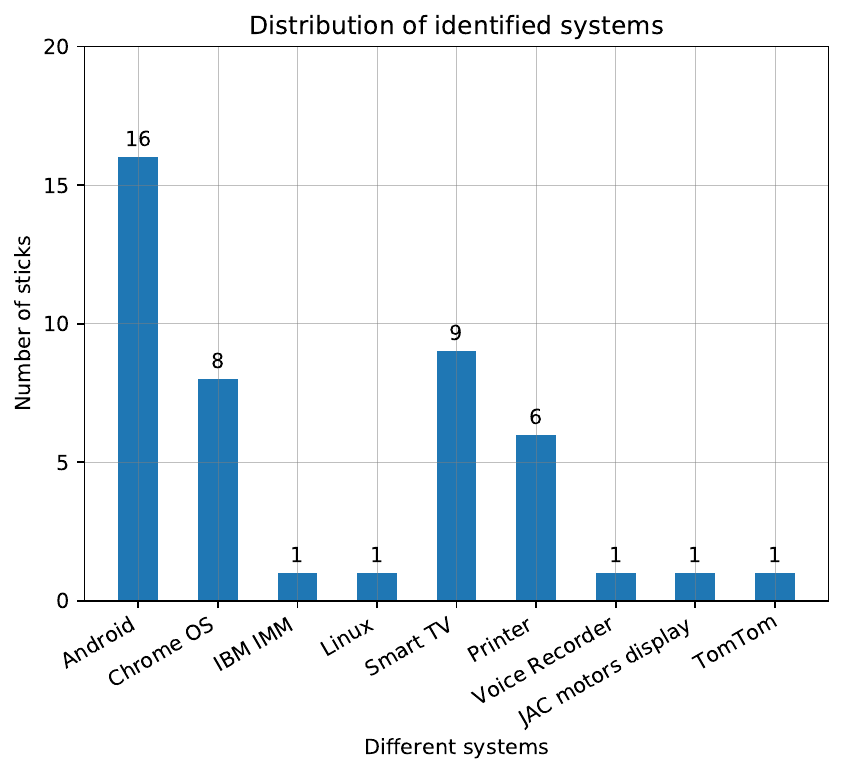}
  \caption{Distribution of different system data found on the USB drives.}
  \label{fig:fau_results_systems}
\end{figure}

In a first step, the found data was used to identify possible devices
or systems the chip originally could have been built in.
This way it can be distinguished whether only the chip was recycled or the whole USB drive was reused.
To achieve this a reverse image search on the found icons, logos, photos, pictures and
wallpapers was performed.  Furthermore, the archive files were
unpacked and analyzed which resulted in much OS and App related data
like operating system settings or application graphics.  Other files
were simply browsed through.  Finally, the files were searched for
certain terms like ``Android'' or ``Chrome'' using \cmd{grep}.
Thereby, three originally used operating systems could be identified:
Android, Chrome OS and Linux.  Furthermore, the previous usage as
smart TV (most of them were Samsung Smart TVs), printer and voice recorder could be proven.
Fig.~\ref{fig:fau_results_systems} summarizes which systems could be
verified by the analysis of the reconstructed data on the USB drives.
Our findings indicate that the USB drives were not reused as a whole but the chips have been recycled.

Fig.~\ref{fig:fau_results_systems} shows that 16 USB drives contained Android related data.
On three of them (ID S11-03, S11-04, S11-15) private pictures, videos and movies could be reconstructed. 
One USB drive (ID S11-04) contained recordings of an Asian news broadcast and parts of a kid's movie. %CCTV and Monster AG
One USB drive (ID S11-15) contained three pictures of an Asian child.
On one USB drive (ID S11-03) we were able to reconstruct 10298 gifs and jpgs, whereby most of them were of private nature including pictures of young Asian (sometimes half naked or naked) women and babies.

This was the largest finding of data of the experiment.
Unfortunately, in all three cases we were not able to analyze any kind of metadata.

One USB drive (ID S3-32) contained a recording of a private conversation in Chinese.
Other data found on this USB drive indicates that the chip could have been part of a Sony speech recorder.

Furthermore, 27 USB drives contained some kind of \\world map, which could be an indication for a navigation system or weather data.
One USB drive (ID S14-05) contained data that implies a former usage as Saregama music box.
Since those possible systems could not clearly be verified, these USB drives
are not contained in Fig.~\ref{fig:fau_results_systems}.

As mentioned in Sec.~\ref{sec:market_players} many USB drive manufacturers source their chips from Taiwan or South Korea.
During the analysis of the ordered USB drives we could observe that many USB drives contained data of Korean origin (Posters and covers of Korean TV shows, Korean voice recordings and pictures of Korean TV and music stars).
Furthermore, on some chips the inscription ``Taiwan'' could be found.
Besides, data from other Asian locations like India or China have been found and some USB drives also bear the lettering ``Japan''.

\subsection{USB Drive and Chip Characteristics}

Besides the analysis of the reconstructed data, we also evaluated
specific characteristics for correlations with the discovery of data.
Correlations to the following properties were investigated:
\begin{itemize}
\item Supplier,
\item NAND technology,
\item chip architecture standard, and 
\item chip manufacturer.
\end{itemize}

\begin{figure}[h!]
    \includegraphics[width=\linewidth]{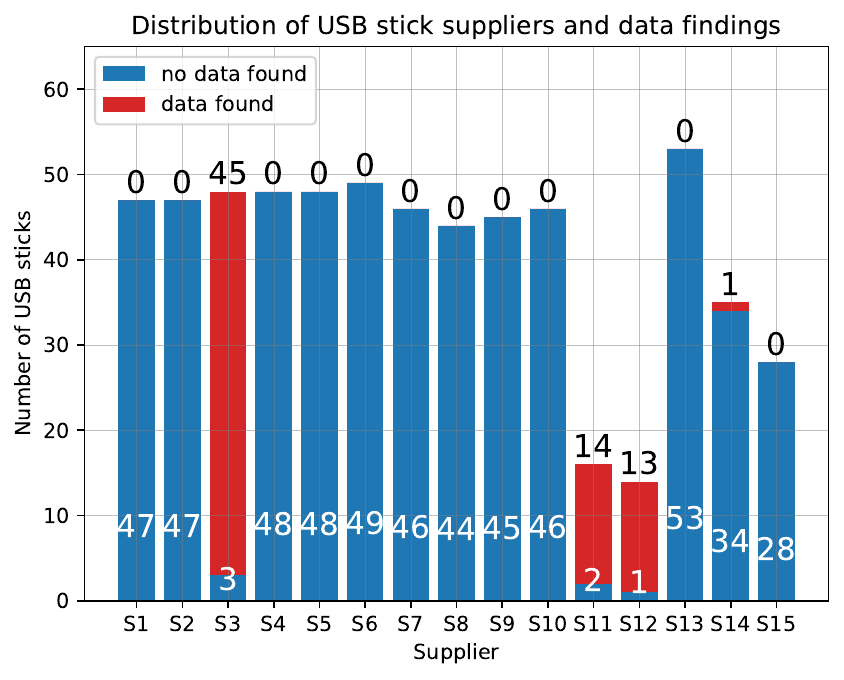}
    \caption{Distribution of USB drive suppliers and data
      findings. Blue indicates that the USB drives did not contain any
      data. Red indicates that the USB drives contained data.}
    \label{fig:FAU_results_suppliers}
\end{figure}

Fig.~\ref{fig:FAU_results_suppliers} shows the distribution of data
findings over the different suppliers. The suppliers have been
anonymized for data protection reasons. From looking at the data, it
is clear that there is a high correlation between data found and
particular suppliers (ID S3, S11 and S12).

\subsection{Visual Inspection}
To investigate if there is a chance to visually determine if a chip could have been reused the analysts were asked to open the USB drive case and perform a visual inspection of the USB drive itself, the circuit board and the NAND chip.
Thereby, the following could be observed:
\begin{itemize}
	\item Regular serial and manufacturer specific numbers and inscriptions
	\item Regular manufacturer logos
	\item Glue and Epoxy
	\item Scratches				%S3-42, S1-34, S10-11, S3-06, S3-14, S3-16, S3-29, S3-30, S5-01, S5-07, S5-15, S1-47, S9-31, S3-15, S5-22, S7-18
	\item Dirt					%S3-49, S3-32, S3-27, S3-34, S7-15, S3-47
	\item Flux
	\item Paint
	\item Irregular stamps		%S9-11, S9-09, S9-28, S6-27, S9-37, S3-05, S9-24, S3-40, S9-48, S9-22, S9-36, S3-23, S3-48, S9-12, S9-26, S9-32, S3-02, S3-08, S9-4, S10-37, S5-27
	\item Handwritten notes		%S5-03, S5-25, S10-32, S5-36, S5-05
\end{itemize}
% 28 stamps, 10 paint, 9 scratches, 8 dirt, 14 flux, 6 handwritten notes
% Davon mit Daten 4 dirt, 7 paint, 8 scratches, 9 stamp

\begin{figure*}
\minipage{0.32\textwidth}
	\includegraphics[width=\linewidth]{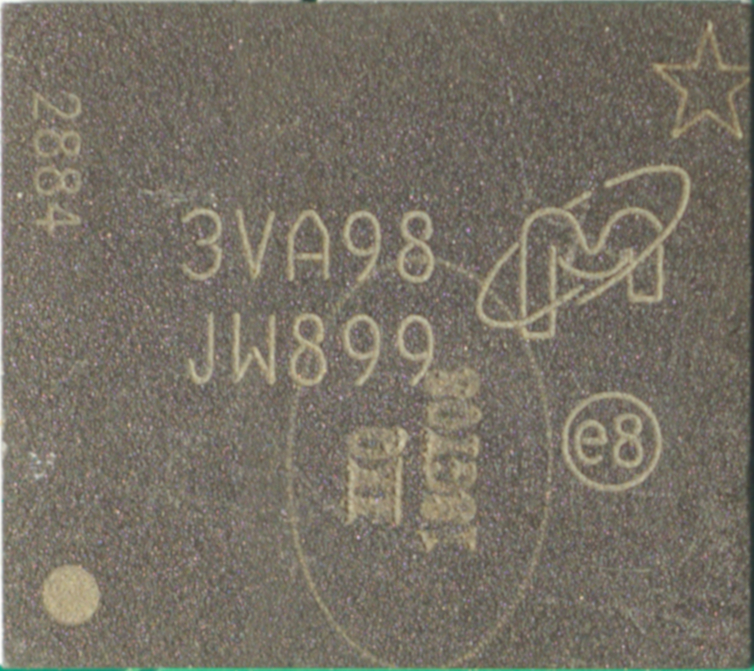}
\endminipage\hfill
\minipage{0.32\textwidth}
	\includegraphics[width=\linewidth]{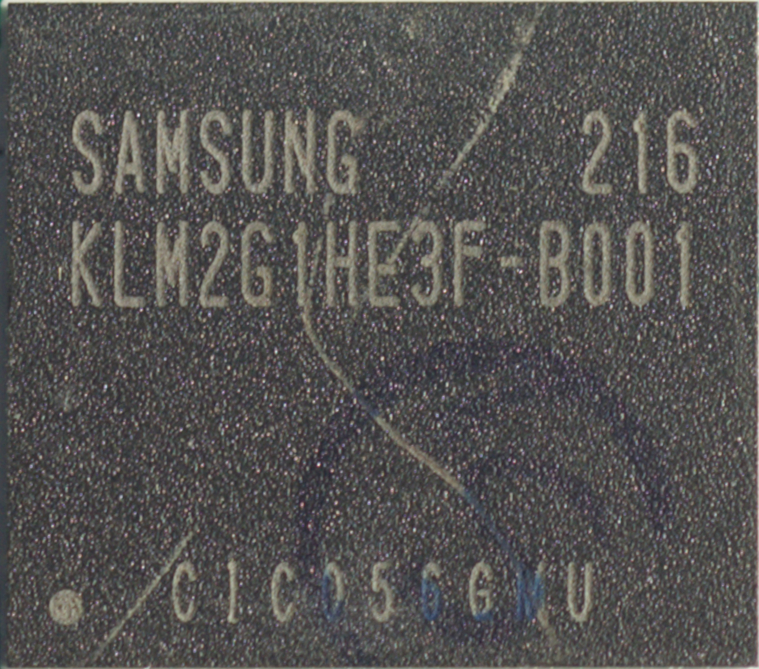}
\endminipage\hfill
\minipage{0.32\textwidth}%
	\includegraphics[width=\linewidth]{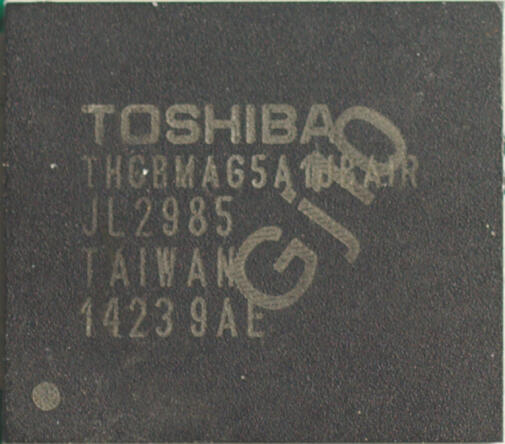}
\endminipage
\caption{Example NAND chips with irregular stamps on it.}
\label{fig:stamps}
\end{figure*}

\begin{figure*}
\minipage{0.32\textwidth}
	\includegraphics[width=\linewidth,height=0.15\textheight]{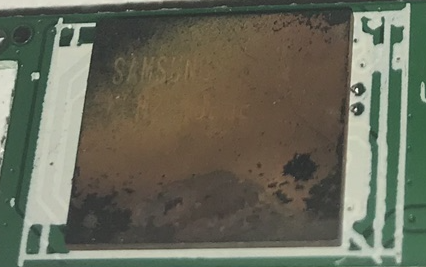}
\endminipage\hfill
\minipage{0.32\textwidth}%
	\includegraphics[width=\linewidth,height=0.15\textheight]{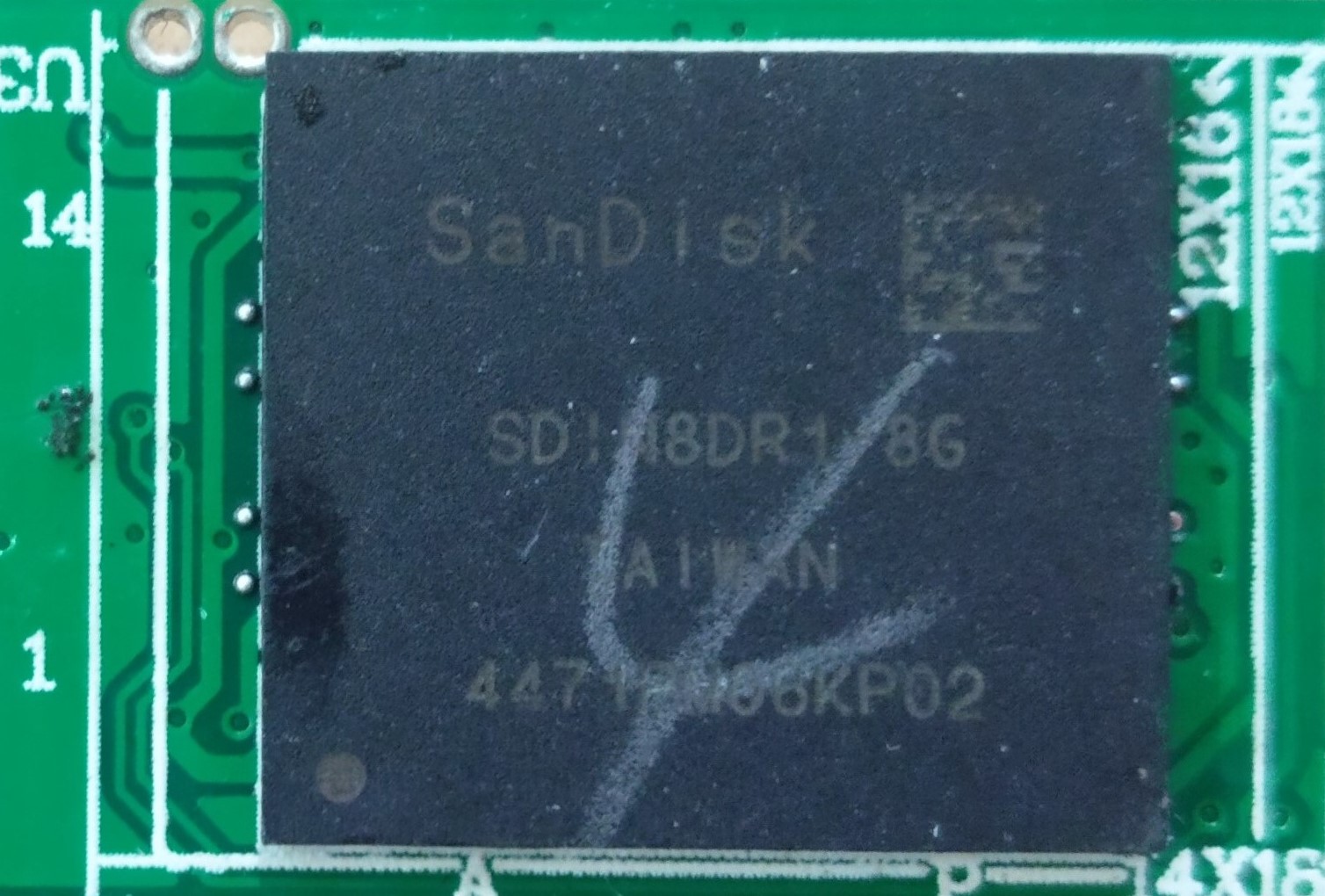}
\endminipage\hfill
\minipage{0.32\textwidth}%
	\includegraphics[width=\linewidth,height=0.15\textheight]{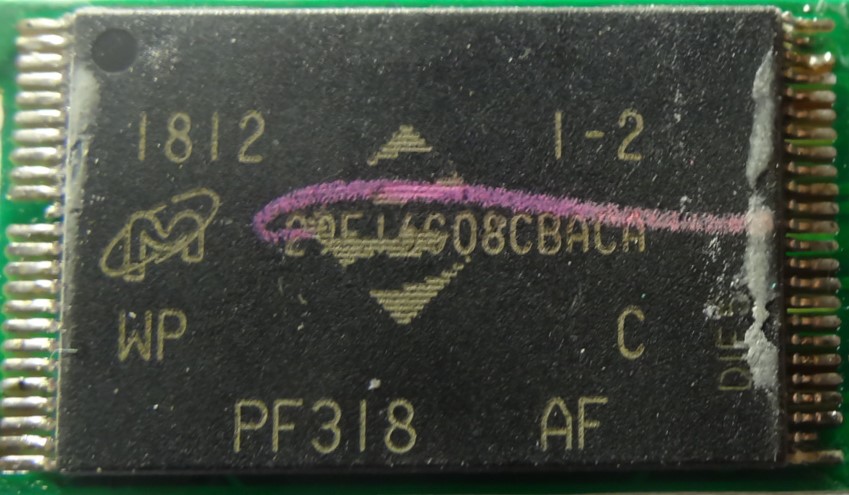}
\endminipage
\caption{Example NAND chips with a handwritten note, dirt and paint.}
\label{fig:deviations}
\end{figure*}

On some chips irregular stamps could be observed.
By comparing these chips with pictures of similar chips the difference between regular and irregular stamps and inscriptions could be established.
According to Westman \cite{Westman:2018:WDT} such stamps are a sign of re-usage of chips, where the stamps are applied to the chip during quality control.
Fig.~\ref{fig:stamps} shows three examples where different irregular stamps could be found.
Overall, irregular stamps could be found on 28 chips, whereby the same stamp could be found on chips from different manufacturers.
Furthermore, we observed striking paint on 10 chips, handwritten notes on 6 chips and markant dirt on 8 chips.
Thereof 4 USB drives with dirt, 7 with paint, 8 with scratches and 9 with stamps contained data
Fig.~\ref{fig:deviations} shows four examples where possible signs of re-use could be found on the chip.

\subsection{Correlation}
\begin{figure*}[t]
    \includegraphics[width=\textwidth]{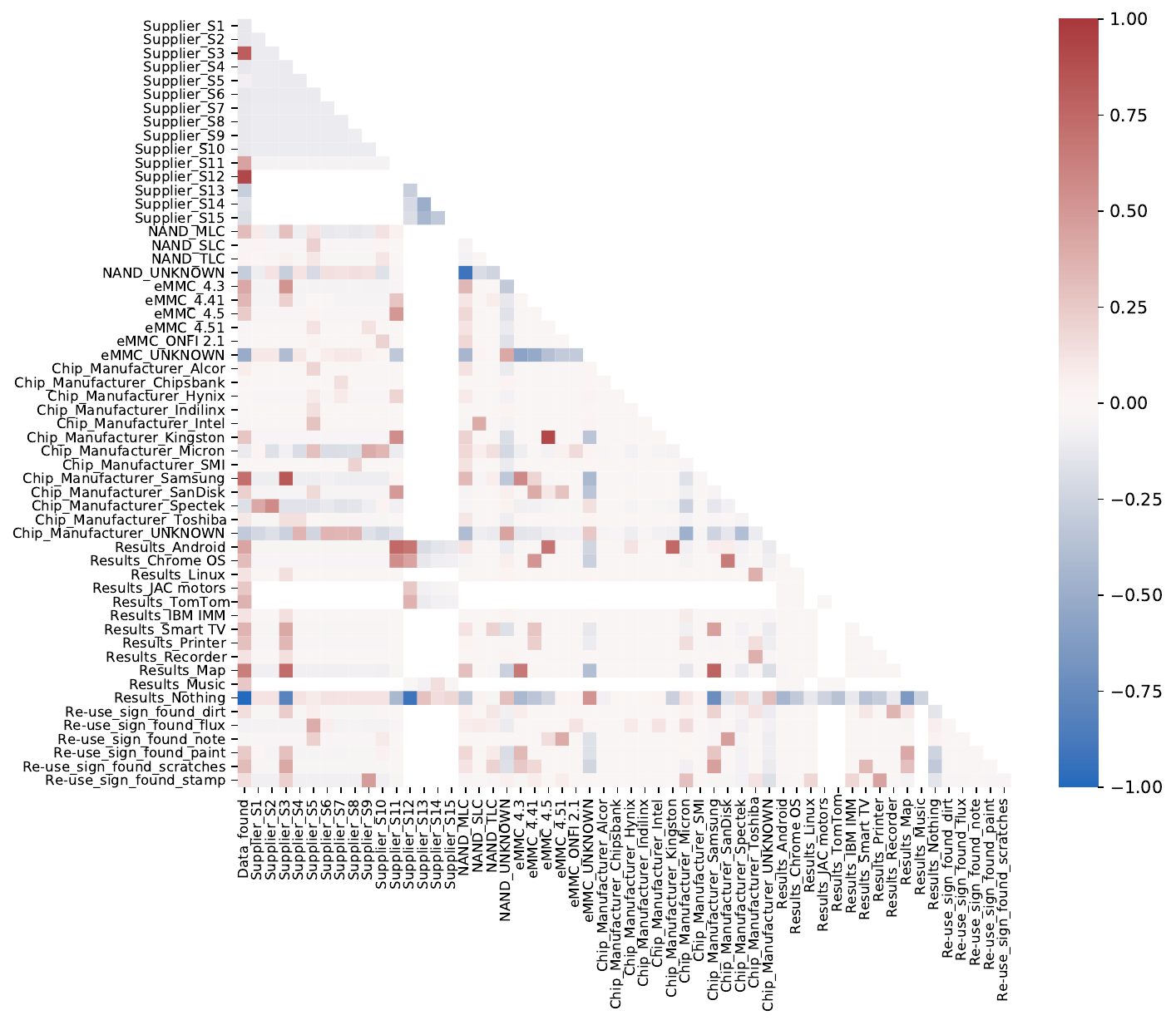}
    \caption{Correlation between different USB drive and chip characteristics and data findings.}
    \label{fig:Results_correlation}
\end{figure*}

For a more complete picture, Fig.~\ref{fig:Results_correlation} shows
the Pearson correlation index for all USB drive and chip
characteristics as heatmap.
This index is a simple way to determine the linear relationship between two variables and indicates the strength of a correlation.
The correlation coefficient by Pearson can range from one to minus one, where one implies a perfectly positive and minus one a perfectly negative correlation.
In Fig.~\ref{fig:Results_correlation} values below zero are depicted in blue and values above zero are depicted as red.
To calculate the correlation, the Python Library Pandas \cite{Pandas} (\texttt{pandas.DataFrame.corr}) was used.
Before the Pearson correlation was calculated the data was normalized by using the One-Hot encoding (\texttt{pandas.\\get\_dummies}).
To generate the heatmap the Python Library Seaborn \cite{Seaborn} (\texttt{seaborn.heatmap}) was used.

At first we used the index to evaluate whether there is a correlation between specific visual characteristics and the finding of non-trivial user data by correlating specific features and the item ``Data found'', which indicates the finding of non-trivial user data.
However, the analysis of the heatmap shows that there are a number of interesting relationships.

For USB drives of the suppliers S12 to S13 no visual inspection was performed. Therefore, the NAND technology, chip architecture, chip manufacturer and visual irregularities could not be evaluated for these drives. This applies to 130 of the 614 analyzed USB drives. Because of this, but also because not all data types could be observed on all USB drives, the correlation matrix contains some empty fields.

The heatmap shows that there is no clear correlation between data
findings and the NAND technology, chip
architecture standard or the signs of re-use.
However, there is a correlation between ``Data
found'' and the chip manufacturer Samsung (Pearson coefficient of
0.716).  Interestingly the Samsung chips do not strongly correlate to
the Smart TV data (0.459) but to the finding of maps (0.777).  The
correlation coefficient between the chip manufacturer Kingston and the
discovery of Android (0.742) could indicate a relationship between
these two characteristics.  Furthermore, there seems to be a
relationship between the chip manufacturer SanDisk and Chrome OS
(0.641).

Overall, the results indicate no clear correlation between any useful
external factors. The correlation with different suppliers indicates
that if one USB drive from a batch contains data, then the probability
is high that other drives from the batch contain data too. Since we
focused on low-cost USB drives where the probability of coming across
a recycled chip is arguably highest, we believe that the probability
of finding data will be lower for more high-end brands that focus on
quality and not on price. To investigate this is part of future work.

\section{Legal Implications}
\label{sec:legal}

Law enforcement agencies and the judiciary face a problem that most of them are not even aware of: Data saved on USB drives (such as photos and documents) can no longer easily be used as ``meaningful'' evidence in criminal proceedings (neither in preliminary proceedings nor in a trial).
As described above, it can happen that old data is found on USB drives purchased as new, of which the owner of the USB drive has no knowledge.
We now discuss the legal implications of this situation, a situation that can be dramatic to the owner of such an USB drive, as the following scenario clarifies (in which we refer to articles of the German Law for brevity and precision): A successful middle-aged public known businessman faces the charge of tax evasion.
As soon as there is an initial suspicion (sufficient factual evidence, § 152 II German Code of Criminal Procedures), criminal tax law proceedings are initiated.
The reason for this can already be given by possible indications of third parties (e.g. disgruntled business partners or competitors) [20, p. 25].
In the course of the criminal tax proceedings, the respective investigating tax authority (if necessary also the public prosecutor’s office, § 386 I, IV German Regulation of Taxation) uses, among other things and under certain conditions, the means of search and seizure (§ 399 I German Regulation of Taxation in conjunction with §§ 102 ff., 98 ff. German Code of Criminal Procedures) - mainly of (business) documents, computers and data carriers, including USB drives.
After the forensic examination of these drives, already deleted files could be recovered, showing, among other things, posing pictures of young Asian girls or naked babies.
It was also possible to inspect "copyright-protected" film clips and pieces of music as well as private addresses and sound recordings of private conversations.
On the basis of these accidental findings (in the meaning of § 108 I German Code of Criminal Procedures), which might indicate the commission of further criminal offenses, such as § 184b German Criminal Code (possession of child pornographic writings), § 201a German Criminal Code (violation of the most personal sphere of life through image recordings), § 106 I German Copyright Law (unauthorized exploitation of copyright protected works), §§ 43, 44 German Federal Data Protection (misuse of personal data on the Internet), the public prosecutor’s office now initiates further investigative measures (§ 160 German Code of Criminal Procedures).
Such investigative measures (e.g. §§ 100a ff. German Code of Criminal Procedures) may impose damaging consequences for the person concerned in his or her reputation as well as severe restrictions in his or her rights of freedom and in his or her personal and intimate sphere.
The attempt to exonerate oneself - the USB drive had been bought new and had never been used before, so the existence of the data could not be explained - will probably "fall on deaf ears" with most of the accused.
The current state of knowledge of the law enforcement authorities and the judiciary is more likely to lead to dismissing this as a weak excuse and assuming that the accused had deleted the data before the seizure in order to destroy possible evidence.
The fact that the forensic investigation cannot determine the date when these files were saved or deleted (because the old file structure and file system entries were overwritten during the recycling process) makes matters worse.
This is precisely when law enforcement agencies and the judiciary need to be made aware during their investigations that the probative force and probative value of files recovered on USB drives as evidence in criminal proceedings can be severely limited.
They can no longer be sure that files found on a USB drive really come from the owner of the drive.
The results of this paper show that there is a certain probability for old files to be found on newly purchased USB flash drives that do not originate from the current owner of this flash drive and that the current owner could not even have known of their existence.
For example, when child pornography material is found, the chain of forensic evidence often presented in the past is no longer sufficient [21].
Up to now, when an increasing number of circumstantial evidence \footnote{An auxiliary fact which influences the probability of the existence of a legal element of the offense, e.g. violence in the case of "rape" or generally the perpetration of the accused ["who... kills"].} is available, the court can come to the conclusion (§ 261 German Code of Criminal Procedures) "that there cannot be so many coincidences" and recognize the material truth as proven and bases its judgment on it.
However, the result of this paper is intended to illustrate how dangerous it can be to base one’s conviction or investigations on a chain of circumstantial evidence whose individual pieces of evidence have no or limited probative evidentiary value in themselves.
To ensure the quality of the evidentiary value, the forensic results must be critically examined and questioned by law enforcement agencies, the judiciary and forensic experts \cite{FAZ:2017}.
This is the only way to ensure that the facts of the case are clarified on a solid factual basis (§ 244 II German Code of Criminal Procedures). 

\section{Conclusions and Future Work}
\label{sec:summary}

We reported on a large-scale experiment of sanitization practices of
USB flash drives from the low-price Chinese market. Overall, we found
a non-negligible probability (12 \%) of finding data on cheap but new
USB drives ordered for promotional products. This is a clear
indication of weak sanitization practices in this market sector,
practices that clearly violate good data protection methods and can
even be classified as fraud in some jurisdictions.

We unfortunately found no clearly correlating factors to the existence
of user data apart from the insight that data findings grouped within
the batches we received from suppliers and not across the
batches. This increases the uncertainty of data provenance if data
without metadata is found on USB drives by file carving software. 

By focusing on cheap drives (``cheapest of the cheapest'', commercial
giveaways, no-name and pirated no-name), we could confirm Westman's
conjecture that chip recycling in this part of the market was a
relevant factor that could be used as an excuse of suspects who had
recently deleted incriminating evidence \cite{Westman:2018:WDT}. By
merely looking at cheap drives, we could not investigate price of the
drive as a correlating factor. Since we are unaware of any reported
findings of data on more quality-oriented brands that use eMMC, a
follow-up study should be undertaken with higher priced USB
drives. Since this will require a substantially higher research
budget, we are still looking for possible sponsors for this
experiment.

\subsection*{Acknowledgments}

We wish to thank all students who supported this experiment,
especially the students from the course on ``Advanced Forensic
Computing'' in the winter term 2018/2019 at FAU. We also thank the
anonymous reviewers for their helpful comments on previous versions of
the paper.  Work was supported by Deutsche Forschungsgemeinschaft
(DFG, German Research Foundation) as part of the Research and Training
Group 2475 ``Cybercrime and Forensic Computing'' (grant number
393541319/GRK2475/1-2019).

%
% author contribution according to https://casrai.org/credit/
\printcredits

\bibliography{references}

@Inbook{Reddy2019,
author = {Reddy, Niranjan},
title = {Solid State Device (SSD) Forensics},
bookTitle = {Practical Cyber Forensics: An Incident-Based Approach to Forensic Investigations},
year = {2019},
publisher = {Apress},
address = {Berkeley, CA},
pages = {379--400},
isbn = {978-1-4842-4460-9},
doi = {10.1007/978-1-4842-4460-9_12},
url = {\url{https://doi.org/10.1007/978-1-4842-4460-9\_12}}
}

@InProceedings{10.1007/978-3-319-46279-0_18,
author = {Singh, Bhupendra and Saharan, Ravi and Somani, Gaurav and Gupta, Gaurav},
editor = {Peterson, Gilbert and Shenoi, Sujeet},
title = {Secure File Deletion for Solid State Drives},
booktitle = {Advances in Digital Forensics XII},
year = {2016},
publisher = {Springer International Publishing},
address = {Cham},
pages = {345--362},
isbn = {978-3-319-46279-0}
}

@article{FUKAMI2017S1,
title  =  {Improving the reliability of chip-off forensic analysis of NAND flash memory devices},
journal  =  {Digital Investigation},
volume  =  {20},
pages  =  {S1 - S11},
year  =  {2017},
note  =  {DFRWS 2017 Europe},
issn  =  {1742-2876},
doi  =  {https://doi.org/10.1016/j.diin.2017.01.011},
url  =  {\url{http://www.sciencedirect.com/science/article/pii/S1742287617300415}},
author  =  {Aya Fukami and Saugata Ghose and Yixin Luo and Yu Cai and Onur Mutlu}
}

@InProceedings{10.1007/978-3-030-05487-8_8,
author = {Vieyra, John and Scanlon, Mark and Le-Khac, Nhien-An},
editor = {Breitinger, Frank and Baggili, Ibrahim},
title = {Solid State Drive Forensics: Where Do We Stand?},
booktitle = {Digital Forensics and Cyber Crime},
year = {2019},
publisher = {Springer International Publishing},
address = {Cham},
pages = {149--164},
isbn = {978-3-030-05487-8}
}

@book{SSD,
author = {Rino Micheloni, Alessia Marelli, Kam Eshghi},
title = {Inside Solid State Drives (SSDs)},
date = {2018},
publisher = {Springer, Singapore},
isbn = {978-981-13-0598-6},
doi = {https://doi.org/10.1007/978-981-13-0599-3},
}

@misc{Seaborn,
author = {Michael Waskom},
title = {seaborn: statistical data visualization — seaborn 0.11.0 documentation},
date = {2020},
url = {\url{https://seaborn.pydata.org/}},
}

@misc{Pandas,
ALTauthor = {author},
ALTeditor = {editor},
title = {pandas - Python Data Analysis Library},
date = {2020},
url = {\url{https://pandas.pydata.org/}},
}

@Misc{FAZ:2017,
  OPTkey = 	 {},
  author = 	 {Peter Welchering},
  title = 	 {{R\"atselhafte Daten auf fabrikneuen USB-Sticks}},
  howpublished = {Frankfurter Allgemeine Zeitung},
  month = 	 {May 17},
  year = 	 {2017},
  note = 	 {\url{http://www.faz.net/aktuell/technik-motor/digital/recycling-neue-usb-sticks-enthalten-oft-restdaten-15015418.html}},
  OPTannote = 	 {}
}

@Misc{Westman:2018:WDT,
  OPTkey = 	 {},
  author = 	 {Martin Westman},
  title = 	 {Where Did That Incriminating Evidence Come From?},
  howpublished = {Workshop at DFRWS EU 2018, Florence, Italy},
  month = 	 {March},
  year = 	 {2018},
  note = 	 {\url{https://dfrws.org/presentation/where-did-that-incriminating-evidence-come-from/}},
  OPTannote = 	 {}
}

@manual{JESD84-A43:2007,
  author = {},
  title = {Embedded Multimediacard (eMMC) eMMC/Card Product Standard, High Capacity, Including Reliable Write, Boot, and Sleep Modes (MMCA, 4.3)},
  organization = {JEDEC SOLID STATE TECHNOLOGY ASSOCIATION},
  address = {3103 North 10th Street, Suite 240 South, Arlington, Virginia 22201-2107},
  edition = {},
  month = {November},
  year = {2007},
  note = {\url{https://www.jedec.org/system/files/docs/JESD84-A43.pdf}}
}

@manual{JESD84-B51:2015,
	author = {},
	title = {Embedded Multi-Media Card (eMMC), Electrical Standard (5.1)},
	organization = {JEDEC SOLID STATE TECHNOLOGY ASSOCIATION},
	address = {3103 North 10th Street, Suite 240 South, Arlington, Virginia 22201-2107},
	edition = {},
	month = {Februar},
	year = {2015},
	note = {\url{https://www.jedec.org/sites/default/files/docs/JESD84-B51.pdf}}
	
}

@manual{ONFI4.2:2020,
	author = {},
	organization = {Intel Corporation, Micron Technology Inc., Phison Electronics Corp., Western Digital Corporation, SK Hynix Inc., Sony Corporation},
        title = {ONFi Specification 4.2},
address = {},
	edition = {},
	month = {Februar},
	year = {2020},
	note = {\url{http://www.onfi.org/specifications}}

}

@inproceedings{DBLP:conf/imf/FreilingHM08,
  author    = {Felix C. Freiling and
               Thorsten Holz and
               Martin Mink},
  title     = {Reconstructing People's Lives: {A} Case Study in Teaching Forensic
               Computing},
  booktitle = {IT-Incidents Management {\&} IT-Forensics - {IMF} 2008, Conference
               Proceedings, September 23-25, 2008, Mannheim, Germany},
  pages     = {125--142},
  year      = {2008},
  OPTcrossref  = {DBLP:conf/imf/2008},
  url       = {\url{http://subs.emis.de/LNI/Proceedings/Proceedings140/article2299.html}},
  bibsource = {dblp computer science bibliography, https://dblp.org}
}

@article{DBLP:journals/dsonline/GarfinkelS03,
  author    = {Simson L. Garfinkel and
               Abhi Shelat},
  title     = {{IEEE} Security {\&} Privacy: Data Forensics - Rememberance of
               Data Passed: {A} Study of Disk Sanitization Practices},
  journal   = {{IEEE} Distributed Systems Online},
  volume    = {4},
  number    = {2},
  year      = {2003},
  bibsource = {dblp computer science bibliography, https://dblp.org}
}

@article{DBLP:journals/ieeesp/GarfinkelS03,
  author    = {Simson L. Garfinkel and
               Abhi Shelat},
  title     = {Remembrance of Data Passed: {A} Study of Disk Sanitization Practices},
  journal   = {{IEEE} Security {\&} Privacy},
  volume    = {1},
  number    = {1},
  pages     = {17--27},
  year      = {2003},
  url       = {\url{https://doi.org/10.1109/MSECP.2003.1176992}},
  doi       = {10.1109/MSECP.2003.1176992},
  bibsource = {dblp computer science bibliography, https://dblp.org}
}

@Book{Proust:2001:ISL,
  author = 	 {Marcel Proust},
  title = 	 {In Search of Lost Time},
  publisher = 	 {Penguin Classics},
  year = 	 {2002},
  OPTkey = 	 {},
  OPTvolume = 	 {},
  OPTnumber = 	 {},
  OPTseries = 	 {},
  OPTaddress = 	 {},
  OPTedition = 	 {},
  OPTmonth = 	 {},
  OPTnote = 	 {},
  OPTannote = 	 {}
}

@inproceedings{DBLP:conf/iciss/WrightKS08,
  author    = {Craig S. Wright and
               Dave Kleiman and
               Shyaam Sundhar R. S.},
  editor    = {R. Sekar and
               Arun K. Pujari},
  title     = {Overwriting Hard Drive Data: The Great Wiping Controversy},
  booktitle = {Information Systems Security, 4th International Conference, {ICISS}
               2008, Hyderabad, India, December 16-20, 2008. Proceedings},
  series    = {Lecture Notes in Computer Science},
  volume    = {5352},
  pages     = {243--257},
  publisher = {Springer},
  year      = {2008},
  url       = {\url{https://doi.org/10.1007/978-3-540-89862-7\_21}},
  doi       = {10.1007/978-3-540-89862-7\_21},
  bibsource = {dblp computer science bibliography, https://dblp.org}
}

@InProceedings{Nisbet:2013:AFA,
  author = 	 {A. Nisbet and S. Lawrence and M. Ruff},
  title = 	 {A forensic analysis and comparison of solid state drive data retention with trim enabled file systems},
  OPTcrossref =  {},
  OPTkey = 	 {},
  booktitle = {11th Australian Digital Forensics Conference},
  year = 	 {2013},
  OPTeditor = 	 {},
  OPTvolume = 	 {},
  OPTnumber = 	 {},
  OPTseries = 	 {},
  pages = 	 {103--111},
  OPTmonth = 	 {},
  OPTaddress = 	 {},
  OPTorganization = {},
  OPTpublisher = {},
  OPTnote = 	 {},
  OPTannote = 	 {}
}

@article{doi:10.1080/1206212X.2017.1289689,
author = {Nikki Robins and Patricia A. H. Williams and Krishnun Sansurooah},
title = {An investigation into remnant data on USB storage devices sold in Australia creating alarming concerns},
journal = {International Journal of Computers and Applications},
volume = {39},
number = {2},
pages = {79-90},
year  = {2017},
publisher = {Taylor & Francis},
doi = {10.1080/1206212X.2017.1289689},
URL = {\url{https://doi.org/10.1080/1206212X.2017.1289689}},
eprint = {https://doi.org/10.1080/1206212X.2017.1289689}
}

@Misc{Heise:2019,
  OPTkey = 	 {},
  author = 	 {Martin Oversohl},
  title = 	 {{Gel\"oscht und doch ausgestellt – Galerie zeigt Fotos aus Festplatten von eBay}},
  howpublished = {\url{https://www.heise.de/newsticker/meldung/Geloescht-und-doch-ausgestellt-Galerie-zeigt-Fotos-aus-Festplatten-von-eBay-4492566.html}},
  month = 	 sep,
  year = 	 {2019},
  OPTnote = 	 {},
  OPTannote = 	 {}
}

@inproceedings{wang2008economic,
  title={Economic conditions for formal and informal recycling of e-waste in China},
  author={Wang, Feng and Huisman, Jaco and Marinelli, Thomas and Zhang, Yuqi and van Ooyen, Stephan},
  booktitle = {International Conference Electronics Goes Green},
  year={2008},
  OPTorganization={Electronics Goes Green}
}

@article{chi2011informal,
    title={Informal electronic waste recycling: a sector review with special focus on China},
    author={Chi, Xinwen and Streicher-Porte, Martin and Wang, Mark YL and Reuter, Markus A},
    journal={Waste Management},
    volume={31},
    number={4},
    pages={731--742},
    year={2011},
    publisher={Elsevier}
}

@Misc{PSI,
  OPTkey = 	 {},
  author = 	 {{International Network of the Promotional Product Industry}},
  title = 	 {PSI Network},
  howpublished = {\url{https://www.psi-network.de/en/}},
  OPTmonth = 	 {},
  year = 	 {2020},
  OPTnote = 	 {},
  OPTannote = 	 {}
}

@inproceedings{DBLP:conf/dfrws/RichardR05,
  author    = {Golden G. Richard III and
               Vassil Roussev},
  title     = {Scalpel: {A} Frugal, High Performance File Carver},
  booktitle = {Refereed Proceedings of the 5th Annual Digital Forensic Research Workshop,
               {DFRWS} 2005, Astor Crowne Plaza, New Orleans, Louisiana, USA, August
               17-19, 2005},
  year      = {2005},
  url       = {\url{http://www.dfrws.org/2005/proceedings/richard\_scalpel.pdf}},
  bibsource = {dblp computer science bibliography, https://dblp.org}
}

@Misc{foremost,
  OPTkey = 	 {},
  author = 	 {Kris Kendall and Jesse Kornblum and Nick Mikus},
  title = 	 {Foremost},
  howpublished = {\url{http://foremost.sourceforge.net/}},
  OPTmonth = 	 {},
  year = 	 {2020},
  OPTnote = 	 {},
  OPTannote = 	 {}
}

@Misc{photorec,
  OPTkey = 	 {},
  author = 	 {Christophe Grenier},
  title = 	 {PhotoRec - CGSecurity},
  howpublished = {\url{https://www.cgsecurity.org/wiki/PhotoRec}},
  OPTmonth = 	 {},
  year = 	 {2020},
  OPTnote = 	 {},
  OPTannote = 	 {}
}

\end{document}